\documentclass[11pt]{book}
\usepackage{Wiley-AuthoringTemplate}
\usepackage{chapterbib}
\usepackage[sectionbib,authoryear]{natbib}
\usepackage[left=2cm,right=2cm,top=2cm,bottom=2cm]{geometry}
\begin{document}

\chapter[Solar Flares and Magnetic Helicity]{Solar Flares and Magnetic Helicity}

\author*[1]{Shin Toriumi}
\author[2,3]{Sung-Hong Park}

\address[1]{\orgdiv{Institute of Space and Astronautical Science},
\orgname{Japan Aerospace Exploration Agency},
      \postcode{252-5210}, \countrypart{Kanagawa}, \city{Sagamihara},
      \street{3-1-1 Yoshinodai, Chuo-ku}, \country{Japan}}%

\address[2]{\orgdiv{Institute for Space-Earth Environmental Research},
\orgname{Nagoya University},
     \postcode{464-8601}, \countrypart{Aichi}, \city{Nagoya},
     \street{Furo-cho, Chikusa-ku}, \country{Japan}}%

\address[3]{\orgdiv{W.W. Hansen Experimental Physics Laboratory},
\orgname{Stanford University},
\postcode{94305}, \countrypart{CA},
     \city{Stanford}, \street{452 Lomita Mall}, \country{USA}}%

\address*{Corresponding Author: Shin Toriumi; \email{toriumi.shin@jaxa.jp}}

\maketitle

\begin{abstract}{}
Solar flares and coronal mass ejections are the largest energy release phenomena in the current solar system. They cause drastic enhancements of electromagnetic waves of various wavelengths and sometimes eject coronal material into the interplanetary space, disturbing the magnetic surroundings of orbiting planets including the Earth. It is generally accepted that solar flares are a phenomenon in which magnetic energy stored in the solar atmosphere above an active region is suddenly released through magnetic reconnection. Therefore, to elucidate the nature of solar flares, it is critical to estimate the complexity of the magnetic field and track its evolution. Magnetic helicity, a measure of the twist of coronal magnetic structures, is thus used to quantify and characterize the complexity of flare-productive active regions. This chapter provides an overview of solar flares and discusses how the different concepts of magnetic helicity are used to understand and predict solar flares.
\end{abstract}

\keywords{Solar flares, Coronal mass ejections, Active regions, Magnetic fields, Magnetic helicity}

\section{Introduction}\label{sec:intro}

\begin{figure}
\begin{center}
\includegraphics[width=0.9\textwidth]{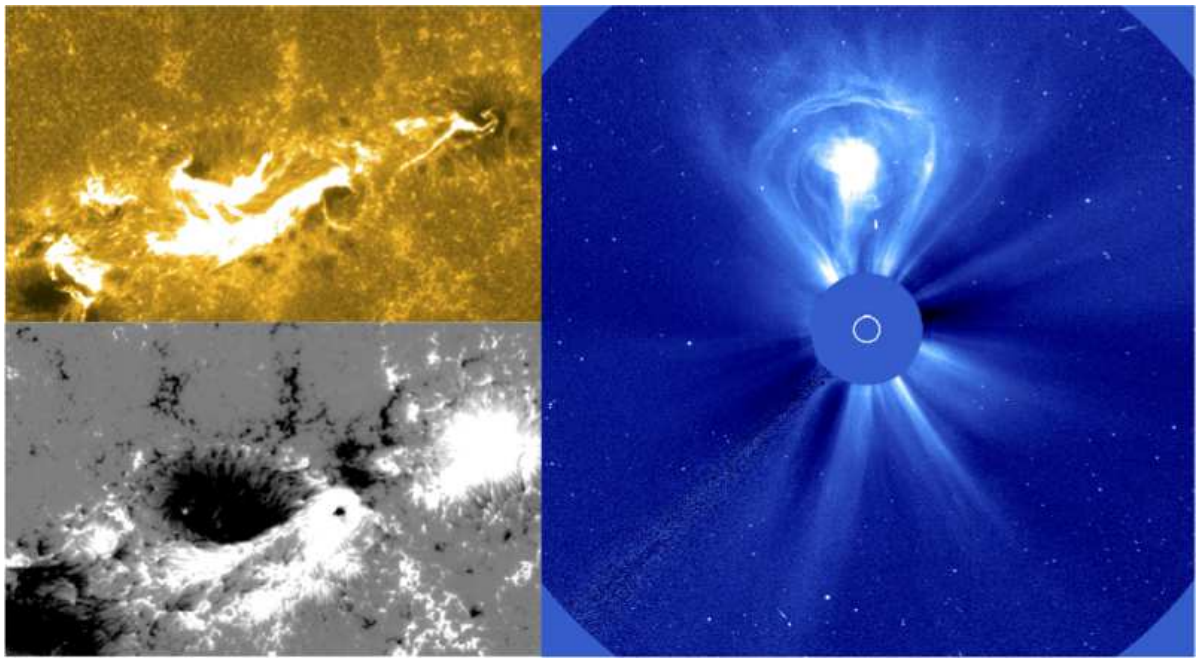}
\end{center}
\caption{Solar flare and CME. (Left) An X2.2-class flare in NOAA AR 11158, as observed by the Hinode satellite. The top panel shows the intensity of the Ca II H line, in which bright ``flare ribbons'' expand across the AR, whereas the bottom panel displays the corresponding photospheric magnetogram (white/black corresponds to positive/negative polarity), both detected by Hinode/SOT \citep{2007SoPh..243....3K,2008SoPh..249..167T}. (Right) Lightbulb-shaped CME captured by SOHO/LASCO on 2000 February 27. Courtesy of SOHO (ESA \& NASA).
\label{fig:flare}}
\end{figure}

In astronomy, the term ``flare'' generally refers to the sudden brightening of electromagnetic waves at a variety of wavelengths. In the Sun, the flares (Figure~\ref{fig:flare}) are occasionally observed as brightenings across the spectrum from X-rays to radio waves with time scales of minutes to hours \citep{2011SSRv..159...19F,2017LRSP...14....2B}. The typical amount of energy released during the flares ranges from $10^{28}$ to $10^{32}$ erg, making flares the largest energy-releasing phenomena in the current solar system.\footnote{The magnitude of solar flares is characterized by the maximum soft X-ray flux of 1-8 {\AA} measured by the Geostationary Operational Environmental Satellite (GOES). The GOES classes are A, B, C, M, and X, corresponding to $10^{-8}$, $10^{-7}$, $10^{-6}$, $10^{-5}$, and $10^{-4}$ W m$^{-2}$ at Earth.} Solar flares affect orbiting planets in different ways. For instance, X-ray and ultraviolet emissions enhance the degree of ionization of planetary atmospheres, accelerating the escape of atmospheres from the planets. Coronal mass ejections (CMEs), often accompanying large-scale flares \citep{2011LRSP....8....1C}, may hit the magnetosphere of the Earth, causing magnetic storms and malfunctions in telecommunication systems. Therefore, in modern civilization with highly advanced information technology, the prediction and forecasting of flares and CMEs is an urgent issue \citep{2016ApJ...829...89B,2021EP&S...73..159K}.

Observations have shown that particularly strong flares are typically associated with active regions (ARs), conglomerates of magnetic flux with sunspots, which strongly suggests that flares are a release of the free magnetic energy ($\Delta E_{\rm mag}$) that is accumulated in the corona during the evolution of ARs:
\begin{equation}
\Delta E_{\rm mag}=E_{\rm mag}-E_{\rm p}
=\int_{V} \frac{\boldsymbol{B}^{2}}{8\pi}\, dV-\int_{V} \frac{\boldsymbol{B}_{\rm p}^{2}}{8\pi}\, dV,
\label{eq:freeenergy}
\end{equation}
where $\boldsymbol{B}$ is the actual magnetic field and $\boldsymbol{B}_{\rm p}$ is the potential (current-free) field, $\nabla\times \boldsymbol{B}_{\rm p}=0$. Here, the potential field can be proven to be the minimum possible magnetic energy for a given distribution of the normal magnetic field $B_{\rm n}$ on the photospheric surface.\footnote{This property is known as Thomson's theorem, which requires the assumption that the magnetic field is solenoidal.} In a modern view, flares occur in association with magnetic reconnection, a magnetohydrodynamic (MHD) process that converts (free) magnetic energy into kinetic energy and thermal energy with nonthermal particle acceleration \citep{2002A&ARv..10..313P,2011LRSP....8....6S}.

Statistically, stronger flares occur in sunspot groups of not only larger spot areas, or larger amounts of magnetic flux in the photospheric surface (${\rm S}_{\rm ph}$),
\begin{equation}
\Phi =\int_{\rm S_{ph}} |B_{\rm n}|\, dS,
\label{eq:magneticflux}
\end{equation}
but also with more complex shapes \citep{2000ApJ...540..583S,2017ApJ...834...56T}. Most of the sunspot groups are categorized as $\beta$-spots, in which umbrae of the positive and negative polarities have their own penumbrae. However, it is known that the strongest flares tend to occur in sunspot groups called $\delta$-spots, in which the positive and negative polarities are close to each other and surrounded by a single penumbra. In such ARs, magnetic structures often display signs of helical configurations, such as sheared polarity inversion lines (PILs), rotating sunspots, and magnetic flux ropes \citep{2019LRSP...16....3T}. Therefore, magnetic helicity is widely used as an index to quantitatively measure the degree of twist and complexity of ARs. It is expected that elucidating the relationship between the magnetic helicity supplied to the solar surface and accumulated magnetic energy will lead to an understanding of flares and CMEs \citep{2010ApJ...718...43P,2012ApJ...750...48P,2012ApJ...759L...4T,2017SoPh..292...66K}, and give rise to better prediction and forecasting capabilities \citep{2014SoPh..289.3549B,2015ApJ...798..135B,2016ApJ...821..127B,2019ApJS..243...36L}.

This chapter discusses the concept of magnetic helicity in the context of AR evolution and its relationship with flare eruptions. For general accounts of flares and CMEs, readers may refer to other reviews by, e.g., \citet{2002A&ARv..10..313P}, \citet{2011SSRv..159...19F}, \citet{2011LRSP....8....6S}, and \citet{2017LRSP...14....2B}. Regarding the observational characteristics of the flare-productive ARs and the attempts to theoretically reproduce them, refer to \citet{2019LRSP...16....3T}.

The remainder of this chapter is organized as follows. In Section \ref{sec:concepts}, we provide an overview of the basic concepts and formulations of magnetic helicity. In Section \ref{sec:methods}, we provide methods for measuring the volume helicity and helicity flux. Then, in Section \ref{sec:applications}, we review some practical applications of these methods to actual AR data. Section \ref{sec:simulations} describes the numerical reconstruction and modeling of coronal magnetic fields, which are important for accurately measuring the helicity and validating the techniques. Finally, in Section \ref{sec:discussion}, we summarize this chapter and discuss future perspectives. We note that, although other forms of helicity (e.g., current helicity and kinetic helicity) are also discussed in the context of flare activity in ARs, in this chapter, we keep our primary focus on magnetic helicity.

\section{Basic concepts and formulations}\label{sec:concepts}

{\it Magnetic helicity} is a well-conserved quantity that represents the topology of a magnetic field, e.g., twists, kinks, and internal linkages \citep{1956RvMP...28..135E,1958PNAS...44..489W,1969JFM....35..117M}. The helicity $H$ of a magnetic field $\boldsymbol{B}$ contained within a given volume $V$ is defined as
\begin{equation}
H=\int_{V} \boldsymbol{A}\cdot \boldsymbol{B}\, dV,
\label{eq:helicity}
\end{equation}
where $\boldsymbol{A}$ denotes the corresponding vector potential of $\boldsymbol{B}$, i.e., $\boldsymbol{B}=\nabla\times \boldsymbol{A}$. It is strictly conserved in ideal MHD \citep{1958PNAS...44..489W}. The dissipation is relatively weak even in resistive MHD \citep{1974PhRvL..33.1139T,1984JFM...147..133B}, including typical flares \citep{1984GApFD..30...79B}, which is one of the reasons why magnetic helicity has been used for studying the eruptivity.

Equation (\ref{eq:helicity}) is gauge-invariant only for magnetic fields fully contained in a closed volume. To circumvent this limitation, for open systems such as the solar corona, \citet{1984JFM...147..133B} and \citet{1985CoPPC...9..111F} introduced the {\it relative helicity},
\begin{equation}
H_{\rm R}=\int_{V} (\boldsymbol{A}+\boldsymbol{A}_{0})\cdot (\boldsymbol{B}-\boldsymbol{B}_{0})\, dV,
\end{equation}
where $\boldsymbol{B}_{0}$ is the reference field and $\boldsymbol{A}_{0}$ is its vector potential, $\boldsymbol{B}_{0}=\nabla\times \boldsymbol{A}_{0}$. For practical purposes, the potential field $\boldsymbol{B}_{\rm p}$ is often chosen as the reference field,
\begin{equation}
H_{\rm R}=\int_{V} (\boldsymbol{A}+\boldsymbol{A}_{\rm p})\cdot (\boldsymbol{B}-\boldsymbol{B}_{\rm p})\, dV.
\label{eq:relativehelicity}
\end{equation}
The relative helicity (\ref{eq:relativehelicity}) can be decomposed into two separately gauge-invariant components \citep{2003and..book..345B}: $H_{\rm R}=H_{\rm j}+2H_{\rm pj}$ with
\begin{equation}
H_{\rm j}=\int_{V} (\boldsymbol{A}-\boldsymbol{A}_{\rm p})\cdot (\boldsymbol{B}-\boldsymbol{B}_{\rm p})\, dV,
\end{equation}
\begin{equation}
H_{\rm pj}=\int_{V} \boldsymbol{A}_{\rm p}\cdot (\boldsymbol{B}-\boldsymbol{B}_{\rm p})\, dV,
\end{equation}
where $H_{\rm j}$ is the helicity of the current-carrying, or non-potential, component of the magnetic field, $\boldsymbol{B}_{\rm j}=\boldsymbol{B}-\boldsymbol{B}_{\rm p}=\nabla\times(\boldsymbol{A}-\boldsymbol{A}_{\rm p})$, and $H_{\rm pj}$ is the {\it mutual helicity} between $\boldsymbol{B}_{\rm j}$ and $\boldsymbol{B}_{\rm p}$.

However, care must be taken when considering Equation (\ref{eq:relativehelicity}) because the integrand  $(\boldsymbol{A}+\boldsymbol{A}_{\rm p})\cdot (\boldsymbol{B}-\boldsymbol{B}_{\rm p})$ is only meaningful if integrated over the whole domain, and the simple sum of the relative helicity in contiguous subvolumes is not equal to the relative helicity of the entire volume \citep{1984JFM...147..133B,2008ApJ...674.1130L}. To overcome this additivity problem, \citet{2008ApJ...674.1130L} proposed dividing the volume into a set of subvolumes whose boundaries between them are magnetic flux surfaces with no magnetic field penetrating them. The {\it additive self helicity} is thus defined as the relative helicity integrated only over the corresponding subvolume with respect to the reference potential field restricted to the subvolume:
\begin{equation}
H_{{\rm R}i}=\int_{V_{i}}(\boldsymbol{A}+\boldsymbol{A}_{{\rm p}i})\cdot (\boldsymbol{B}-\boldsymbol{B}_{{\rm p}i})\, dV,
\label{eq:additiveselfhelicity1}
\end{equation}
where $V_{i}$ is the subvolume bounded by the flux surface, and $\boldsymbol{B}_{{\rm p}i}$ and $\boldsymbol{A}_{{\rm p}i}$ are its potential field and the vector potential, respectively. The additive self helicity is integrated to obtain the relative helicity:
\begin{equation}
H_{\rm R}=\sum_{i}H_{{\rm R}i},
\label{eq:additiveselfhelicity2}
\end{equation}
where the $i$ indexes denote the subvolumes.

However, in the definition by \citet{2008ApJ...674.1130L}, it is difficult to find a suitable set of subvolumes in practical situations, and the finest possible decomposition may be to take an infinitesimally thin tube around each field line. In this limit, the reference potential field would be the magnetic field itself, so that the additive self helicity would vanish. \citet{2018JPlPh..84f7702Y} followed a different approach, decomposing the relative helicity into a {\it relative field-line helicity,} which vanishes along any field line where the original field matches the reference (potential) field. The integral of the relative field-line helicity gives the relative helicity. Although this is also true for the additive self helicity of Equations (\ref{eq:additiveselfhelicity1}) and (\ref{eq:additiveselfhelicity2}), it is only in relative to the sum of the local reference fields rather than the global reference field used in the relative helicity.

As described in detail in \citet{2015A&A...580A.128P}, the temporal variation in the relative magnetic helicity $H_{\rm R}$ in a closed volume $V$ can be expressed as 
\begin{equation}
\frac{dH_{\rm R}}{dt}=2\int_{V} \boldsymbol{A} \cdot \frac{\partial \boldsymbol{B}}{\partial t}\, dV + \int_{\partial V}\left[(\boldsymbol{A}-\boldsymbol{A}_{\rm p}) \times \frac{\partial(\boldsymbol{A}-\boldsymbol{A}_{\rm p})}{\partial t} \right]\cdot d\boldsymbol{S} -2\int_{V} \boldsymbol{A}_{\rm p} \cdot \frac{\partial \boldsymbol{B}_{\rm p}}{\partial t}\, dV.
\label{eq:dhdt_original}
\end{equation}
It should be noted that only the sum of the three integrals on the right-hand side is gauge-invariant, whereas each integral is not. By using the Faraday's law and the Gauss divergence theorem and assuming an ideal MHD condition at the boundary of the volume (i.e., $\boldsymbol{E}\vert_{\partial V} = (-\boldsymbol{v} \times \boldsymbol{B})\vert_{\partial V}$), $dH_{\rm R}/dt$ of Equation (\ref{eq:dhdt_original}) can be decomposed into two volume integrals and four flux terms on the surface of $V$ as
\begin{equation}
\frac{dH_{\rm R}}{dt}=\frac{dH_{\rm R}}{dt}\bigg\vert_{\rm diss}+ \frac{dH_{\rm R}}{dt}\bigg\vert_{\rm B_{p},var}+ F_{\rm v_{n}}+ F_{\rm B_{n}}+ F_{\rm AA_{p}}+ F_{\rm \phi},
\label{eq:dhdt_original_decomp}
\end{equation}
where
\begin{align}
&\frac{dH_{\rm R}}{dt}\bigg\vert_{\rm diss}= -2\int_{V} \boldsymbol{E} \cdot \boldsymbol{B} \, dV \label{eq:dhdt_original_decomp1_diss} \\
&\frac{dH_{\rm R}}{dt}\bigg\vert_{\rm B_{p},var}= 2\int_{V} \frac{\partial \boldsymbol{\phi}}{\partial t} \nabla \cdot \boldsymbol{A_{\rm p}} \, dV \label{eq:dhdt_original_decomp2_Bp_variation} \\
&F_{\rm v_{n}}= -2\int_{\partial V} (\boldsymbol{B} \cdot \boldsymbol{A}) \boldsymbol{v} \cdot d\boldsymbol{S} \label{eq:dhdt_original_decomp3_flux_normal_v} \\
&F_{\rm B_{n}}= 2\int_{\partial V} (\boldsymbol{v} \cdot \boldsymbol{A}) \boldsymbol{B} \cdot d\boldsymbol{S} \label{eq:dhdt_original_decomp4_flux_normal_B} \\
&F_{\rm AA_{p}}= \int_{\partial V}\left[(\boldsymbol{A}-\boldsymbol{A}_{\rm p}) \times \frac{\partial(\boldsymbol{A}-\boldsymbol{A}_{\rm p})}{\partial t} \right] \cdot d\boldsymbol{S} \label{eq:dhdt_original_decomp5_flux_A_Ap} \\
&F_{\rm \phi}= -2\int_{\partial V} \frac{\partial \boldsymbol{\phi}}{\partial t} \boldsymbol{A}_{\rm p} \cdot d\boldsymbol{S}, \label{eq:dhdt_original_decomp6_flux_Bp}
\end{align}
and $\phi$ is the solution of the Laplace equation $\nabla^{2} \phi = 0$ (i.e., satisfying $\boldsymbol{B}_{\rm p} = \nabla \phi$). Choosing the Coulomb gauge (CG; i.e., $\nabla \cdot \boldsymbol{A}_{\rm p}=0$) for the potential field $\boldsymbol{B}_{\rm p}$, as well as having the following two additional constraints for the boundary condition:
\begin{align}
&\boldsymbol{A}\vert_{\partial V} = \boldsymbol{A}_{\rm p}\vert_{\partial V} \label{eq:dhdt_CG_BC1}, \\
&\boldsymbol{A}_{\rm p} \cdot {\rm d}\boldsymbol{S} \vert_{\partial V} = 0 \label{eq:dhdt_CG_BC2},
\end{align}
we can obtain a simplified expression of $dH_{\rm R}/dt$ as
\begin{equation}
\frac{dH_{\rm R}}{dt}\bigg\vert_{\rm CG}=\frac{dH_{\rm R}}{dt}\bigg\vert_{\rm diss}+ F_{\rm v_{n}}+ F_{\rm B_{n}}.
\label{eq:dhdt_simple}
\end{equation}

\begin{figure}
\begin{center}
\includegraphics[width=\textwidth]{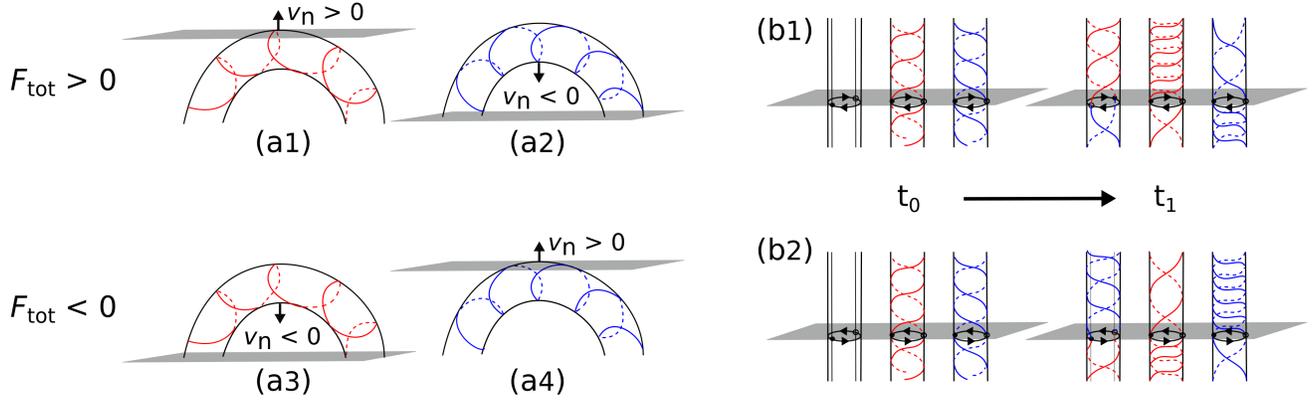}
\end{center}
\caption{Schematic diagram that demonstrates relevant examples of $F_{\rm tot}>0$ (top) and $F_{\rm tot}<0$ (bottom) across the solar surface (shown in gray), in the context of the two distinct helicity flux components $F_{\rm v_{n}}$ (panels a1--a4) and $F_{\rm B_{n}}$ (panels b1--b2). The magnetic flux tubes are described by field lines with right-handed (positive; red) and left-handed (negative; blue) helicity. Here, $t_{0}$ and $t_{1}$ are arbitrarily chosen only to demonstrate the time progress ($t_{0}<t_{1}$).
\label{fig:dhdt_diagram}}
\end{figure}

Now in the context of investigating the transport rate of the relative magnetic helicity across ${\rm S}_{\rm ph}$ of a given AR into the corona, the concept of the {\it relative magnetic helicity flux} $F_{\rm tot}$ has been used in many previous studies of flaring ARs with photospheric magnetic field observations \citep[e.g.,][]{1984JFM...147..133B},
\begin{equation}
\begin{aligned}
F_{\rm tot} &= -(F_{\rm v_{n}}+ F_{\rm B_{n}}) \\
        &=2\int_{\rm S_{ph}} \left[(\boldsymbol{A}_{\rm p}\cdot \boldsymbol{B}_{\rm h})v_{\rm n}-(\boldsymbol{A}_{\rm p}\cdot\boldsymbol{v}_{\rm h})B_{\rm n}\right]\, dS,
\end{aligned}
\label{eq:helicityflux_simple}
\end{equation}
where the subscripts ``n'' and ``h'' indicate the normal and horizontal components, respectively. In Equation (\ref{eq:helicityflux_simple}), the $F_{\rm v_{n}}$ term is associated with the emergence or submergence of a twisted magnetic structure through ${\rm S}_{\rm ph}$ (often called the {\it emergence term}). The $F_{\rm B_{n}}$ term is related to the twisting or untwisting of magnetic field lines by horizontal footpoint motions on ${\rm S}_{\rm ph}$ (the so-called {\it shear term}). The negative sign in front of the sum of $F_{\rm v_{n}}$ and $F_{\rm B_{n}}$ is because the surface unit vector $d\boldsymbol{S}$ is chosen outwardly from the volume $V$ in this particular case of helicity injection into the corona. It should be noted, however, that $F_{\rm tot}$ can be attributed to the transport of helicity into either the coronal domain or the subsurface across ${\rm S}_{\rm ph}$. In Figure~\ref{fig:dhdt_diagram}, the general concept of $F_{\rm tot}$ is demonstrated with notable examples of $F_{\rm tot}>0$ (top) and $F_{\rm tot}<0$ (bottom) by the two distinct helicity flux terms of $F_{\rm v_{n}}$ (panels a1--a4) and $F_{\rm B_{n}}$ (panels b1--b2), respectively.

Once $F_{\rm tot}$ is determined as a function of time, the change in $H_{\rm R}$ due to helicity transport across ${\rm S}_{\rm ph}$ over the time interval $\Delta t = t - t_{0}$ can be obtained:    
\begin{equation}
\Delta H_{\rm R}(t) = \int_{t_{0}}^{t} F_{\rm tot} \,dt,
\label{eq:dh_simple}
\end{equation}
where $t_{0}$ is the reference time (often considered as the start time of observations), and $t$ is the time of interest. In the literature, $\Delta H_{\rm R}$ has often been referred to as the {\it injection} or {\it accumulation} of helicity into the AR coronal volume through ${\rm S}_{\rm ph}$. Nevertheless, strictly speaking, a more appropriate term may be helicity {\it change} as a result of the helicity flux transport. Again, as described in Figure~\ref{fig:dhdt_diagram}, the transport can be achieved not only by injection into the coronal volume but also by removal from the corona into the subsurface.

\section{Methods of helicity estimation}\label{sec:methods}

\subsection{Volume helicity}\label{subsec:methods_volumehelicity}

If the coronal magnetic field is obtained as a result of extrapolation based on photospheric magnetic field measurement or numerical simulation (Section \ref{sec:simulations}), it is possible to directly calculate the helicity in a coronal volume. As mentioned in Section \ref{sec:concepts}, the relative helicity, $H_{\rm R}=\int (\boldsymbol{A}+\boldsymbol{A}_{\rm p})\cdot (\boldsymbol{B}-\boldsymbol{B}_{\rm p})\, dV,$ has a degree of freedom in the choice of gauge. In many practical applications, the relative helicity is calculated for the magnetic field $\boldsymbol{B}=\nabla\times\boldsymbol{A}$ in a finite volume, often of the 3D computational box with the Cartesian coordinates $(x, y, z)$ above a vector magnetic field on the photospheric surface $(z=0)$. In the past, the CG (i.e., $\nabla\cdot\boldsymbol{A}=0$) has been used to calculate the volume helicity \citep[e.g.,][]{2011SoPh..270..165R,2011SoPh..272..243T}. However, the CG, when implemented, could be computationally expensive.

To overcome this issue, \citet{2012SoPh..278..347V} adopted a gauge by \citet{2000ApJ...539..944D}, in which one component of the vector potential is 0, for instance, $A_{z}=0$, which reduces the calculation cost. For a Cartesian box $V=[x_{1}, x_{2}]\times [y_{1}, y_{2}]\times [z_{1}, z_{2}]$, we are free to select a gauge $\hat{\boldsymbol{z}}\cdot\boldsymbol{A}=0$ that satisfies
\begin{equation}
\boldsymbol{A}=\boldsymbol{A}_{0}
-\hat{\boldsymbol{z}}\times \int_{z_{1}}^{z_{2}} \boldsymbol{B}(x, y, z')\, dz',
\end{equation}
where the vector potential $\boldsymbol{A}_{0}$ is a solution to the $z$-component of $\boldsymbol{B}=\nabla\times\boldsymbol{A}$ and is described as $\boldsymbol{A}_{0}=\boldsymbol{A}(x, y, z_{1})=(A_{0x}, A_{0y}, 0)$. We take a simple solution to this:
\begin{align}
&A_{0x} = -\frac{1}{2}\int_{y_{1}}^{y} B_{z}(x, y', z=z_{1})\, dy',\\
&A_{0y} = \frac{1}{2}\int_{x_{1}}^{x} B_{z}(x', y, z=z_{1})\, dx'.
\end{align}
Similarly, the vector potential for the reference potential field, $\boldsymbol{A}_{\rm p}$, is calculated using the same $\boldsymbol{A}_{0}$ as
\begin{equation}
\boldsymbol{A}_{\rm p}=\boldsymbol{A}_{0}
-\hat{\boldsymbol{z}}\times \int_{z_{1}}^{z_{2}} \boldsymbol{B}_{\rm p}(x, y, z')\, dz'.
\end{equation}
The relative helicity for the finite volume, $H_{\rm R}$, can be computed using the set of $\boldsymbol{B}$, $\boldsymbol{A}$, $\boldsymbol{B}_{\rm p}$, and $\boldsymbol{A}_{\rm p}$.

It should be noted here that the calculation results of the relative helicity may differ depending not only on the choice of gauge (CG or DeVore gauge) but also on the implementation method. \citet{2016SSRv..201..147V} systematically compared the relative helicity with multiple calculation methodologies and investigated the difference in calculation results and the dependence on spatial resolution. As a result, it was found, for instance, that the accuracy in computing $H_{\rm R}$ is found to vary by different methods, especially for the CG methods, and that the spread in $H_{\rm R}$ estimates between different methods converge in general at higher spatial resolutions.

\subsection{Magnetic helicity flux}\label{subsec:methods_helicityflux}

To determine the relative magnetic helicity in a coronal volume, it is essential to have a 3D coronal magnetic field, which requires complicated modeling efforts based on either the photospheric magnetic field observation or the polarization of coronal lines. On the other hand, it is straightforward to determine the time series of the magnetic helicity flux $F_{\rm tot}$ and examine the temporal variation and total amount of helicity transported across the AR photosphere.

\begin{figure}
\begin{center}
\includegraphics[width=0.8\textwidth]{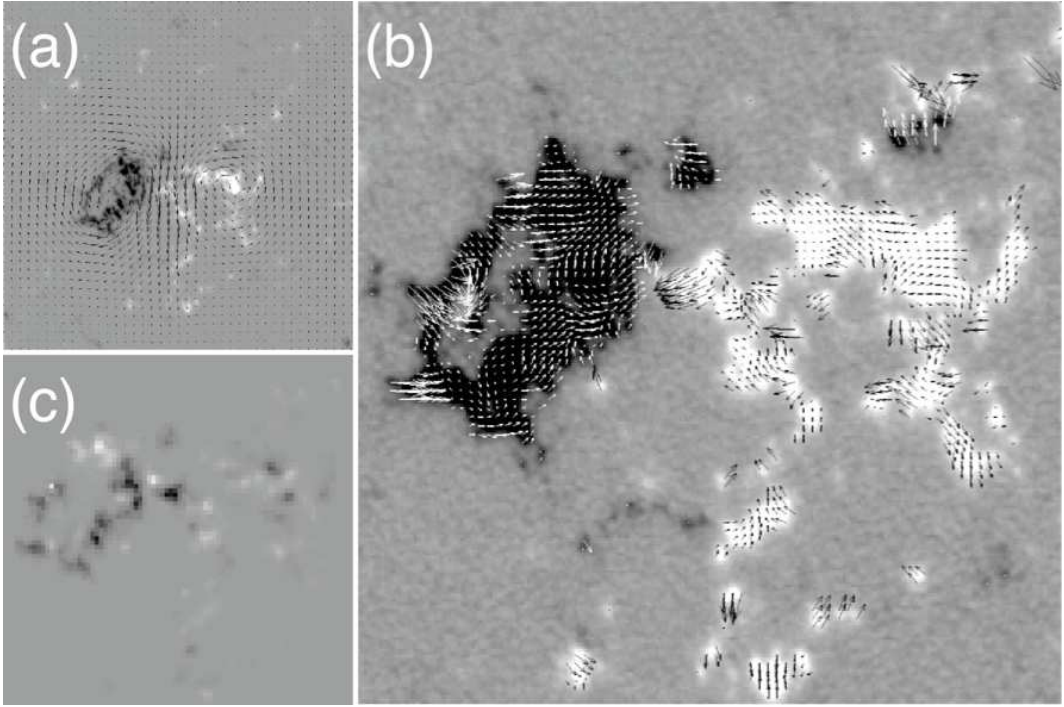}
\end{center}
\caption{An example of estimating $F_{\rm tot}$ with the $F_{\rm tot}^{\rm JC}$ method for NOAA AR 8011. The line-of-sight magnetic field $B_{\rm LOS}$ was obtained by SOHO/MDI when the AR was located near the center of the solar disk at 17:00 UT on 1997 January 17, therefore, it can be well approximated to the normal component $B_{\rm n}$. The grayscale image of $B_{\rm LOS}$ is overplotted with $\boldsymbol{A}_{\rm p}$ (arrows in panel a) and $\boldsymbol{v}_{\rm LCT}$ (arrows in panel b). $F_{\rm tot}$ is estimated by integrating proxy values, i.e., $-2 (\boldsymbol{A}_{\rm p}\cdot\boldsymbol{v}_{\rm LCT})B_{\rm n}$ (grayscale image in panel c), for the relative magnetic helicity flux density over the entire AR photospheric surface under consideration. Figure reproduced from \citet{2001ApJ...560L..95C} by permission of the AAS.
\label{fig:chae2001}}
\end{figure}

One of the first practical methods for estimating $F_{\rm tot}$ in ARs was developed by \citet[][hereafter $F_{\rm tot}^{\rm JC}$]{2001ApJ...560L..95C}, using line-of-sight magnetograms obtained by the Michelson Doppler Imager \citep[MDI;][]{1995SoPh..162..129S} onboard the Solar and Heliospheric Observatory (SOHO),
\begin{equation}
F_{\rm tot}^{\rm JC} = -2\int_{\rm S_{ph}} (\boldsymbol{A}_{\rm p}\cdot\boldsymbol{v}_{\rm LCT})B_{\rm n}\, dS,
\label{eq:chae_helicityflux}
\end{equation}
where $\boldsymbol{v}_{\rm LCT}$ is the horizontal velocity of the {\it apparent} motions of magnetic elements in the photosphere, which is determined using the technique of local correlation tracking \citep[LCT;][]{1988ApJ...333..427N}. $\boldsymbol{A}_{\rm p}$ is calculated from $B_{\rm n}$ using the fast Fourier transform (FFT) as
\begin{equation}
\begin{aligned}
&\boldsymbol{A}_{\rm p,x} = {\rm FFT^{-1}} \left( \frac{jk_{\rm y}}{k_{\rm x}^{2}+k_{\rm y}^{2}} {\rm FFT}(B_{\rm n}) \right), \\
&\boldsymbol{A}_{\rm p,y} = {\rm FFT^{-1}} \left( -\frac{jk_{\rm x}}{k_{\rm x}^{2}+k_{\rm y}^{2}} {\rm FFT}(B_{\rm n})\right).
\end{aligned}
\label{eq:chae_helicityflux_Ap}
\end{equation}
$F_{\rm tot}^{\rm JC}$ is a good proxy of $F_{\rm tot}$ under the condition that $\boldsymbol{v}_{\rm LCT}$ reasonably represents the magnetic field-line footpoint velocity (sometimes called the flux transport velocity) $\boldsymbol{u} \equiv \boldsymbol{v}_{\rm t} - (v_{\rm n}/B_{\rm n}) \boldsymbol{B}_{\rm t}$. However, we caution here that $\boldsymbol{v}_{\rm LCT}$ is not valid in places where newly emerging or completely submerging flux patches appear. Figure~\ref{fig:chae2001} shows an example of the implementation of the $F_{\rm tot}^{\rm JC}$ method for NOAA AR 8011 observed by SOHO/MDI on 1997 January 17.

Another method of estimating $F_{\rm tot}$ was developed by \citet[][hereafter $F_{\rm tot}^{\rm KK}$]{2002ApJ...577..501K}, applying the LCT technique and an inversion from the normal component of the magnetic induction equation, respectively, to infer $\boldsymbol{v}_{\rm h}$ and $v_{\rm n}$, and using the helicity flux formula of Equation (\ref{eq:helicityflux_simple}). Several follow-up studies were carried out using the same $F_{\rm tot}^{\rm KK}$ method, in particular, considering the temporal variations of $F_{\rm v_{n}}$ and $F_{\rm B_{n}}$ separately \citep[e.g.,][]{2003AdSpR..32.1917K,2003AdSpR..32.1949Y}. Since then, $F_{\rm tot}$ has been estimated using several different methods to infer the plasma flow velocity from a sequence of (vector) magnetic field data \citep[e.g.,][]{2004ApJ...610.1148W,2004ApJ...612.1181L,2006ApJ...646.1358S,2008ApJ...683.1134S,2008ApJ...689..593C}.

In addition, there was an approach by \citet[][hereafter $F_{\rm tot}^{\rm EP}$]{2005A&A...439.1191P} to determine $F_{\rm tot}$ explicitly using a specific form of $\boldsymbol{A}_{\rm p}$ that satisfies the CG (i.e.,  $\nabla \cdot \boldsymbol{A}_{\rm p}=0$), 
\begin{equation}
\boldsymbol{A}_{\rm p} = \frac{1}{2\pi} \boldsymbol{\hat{n}} \times \int_{\rm S_{ph}^{\prime}} B_{\rm n}(x^{\prime})\frac{\boldsymbol{r}}{r^{2}}\,dS^{\prime}.
\label{eq:pariat_helicityflux_Ap}
\end{equation}
By inserting $\boldsymbol{A}_{\rm p}$ of Equation (\ref{eq:pariat_helicityflux_Ap}) into Equation (\ref{eq:helicityflux_simple}), $F_{\rm tot}^{\rm EP}$ can be expressed as
\begin{equation}
F_{\rm tot}^{\rm EP} = -\frac{1}{2\pi} \int_{\rm S_{ph}} \int_{\rm S_{ph}^{\prime}} \frac{\boldsymbol{r} \times \left[\boldsymbol{u}-\boldsymbol{u^{\prime}}\right]}{r^{2}} \bigg\vert_{\rm n} B_{\rm n}\,B_{\rm n}^{\prime}\,dS\,dS^{\prime}.
\label{eq:pariat_helicityflux}
\end{equation}
As discussed in \citet{2005A&A...439.1191P}, the $F_{\rm tot}^{\rm EP}$ method was shown to be more efficient in reducing spurious signals in the map of the relative magnetic helicity flux density (i.e., the integrand in Equation (\ref{eq:pariat_helicityflux})) compared to that from the $F_{\rm tot}^{\rm JC}$ method (i.e., the integrand in Equation (\ref{eq:chae_helicityflux})).

\section{Practical applications}\label{sec:applications}

\subsection{Temporal variation}\label{subsec:applications_tempvar}

Since several practical methods were developed in the 2000s to estimate $F_{\rm tot}$ across the AR photosphere from solar magnetic field observations (refer to Section~\ref{subsec:methods_helicityflux}), there have been many studies to find out any temporal variations in the magnetic helicity over various timescales in relation to flaring activity of the source ARs, particularly for comparing differences of flaring versus non-flaring ARs. In the context of timescales under consideration, time series analyses of $F_{\rm tot}$ can be generally classified into two categories: (1) short-term (tens of minutes to a few hours) and (2) long-term (a few hours to several days). In the case of short-term variation, for example, there were reports of a large, impulsive variation in the time series of $F_{\rm tot}$ over the course of the flare \citep[e.g.,][]{2002ApJ...574.1066M,2002ApJ...580..528M,2004SoPh..225..311H,2005A&A...433..683R,2018ApJ...865..139B}. The impulsive variation in $F_{\rm tot}$ typically occurs during the rise phase of the flare X-ray flux, and the magnitude of the impulsive helicity variation (i.e., $\Delta H = \int_{t_{0}}^{t_{1}} F_{\rm tot} \,dt$ with $t_{0}$ and $t_{1}$ as the flare start and end times) correlates with the flare X-ray peak flux. Although this impulsive variation in $F_{\rm tot}$ was found for multiple flare events, we should be cautious because photospheric magnetograms derived from spectropolarimetric observations (in particular, pixels where white-light flare kernels are located) can be significantly affected by flare emission, as mentioned by \citet{2004SoPh..225..311H}.\footnote{An example of such disturbed polarimetric signals can be found in the magnetogram of Figure \ref{fig:flare} of this chapter, in which the core of the positive (negative) polarity sunspot shows a negative (positive) signal.} We note that a detailed investigation of short-term variations in $F_{\rm tot}$ may help us to understand the flare-trigger process (e.g., tether-cutting magnetic reconnection, rapidly growing MHD instabilities), and the relevant flare-associated evolution of the photospheric magnetic field.

\begin{figure}[t!]
\begin{center}
\includegraphics[width=0.95\textwidth]{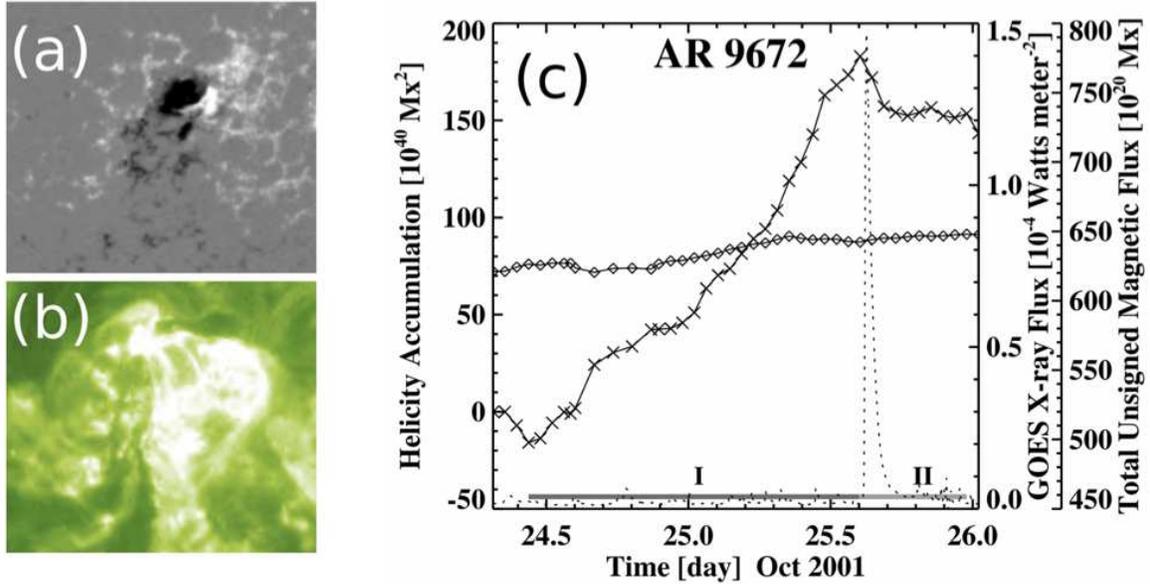}
\end{center}
\caption{Long-term trend of increase in $\Delta H_{\rm R}$ across the photospheric surface of NOAA AR 9672 before the onset of an X1.3 flare on 2001 October 25. The MDI line-of-sight magnetic field and EIT 195\,{\AA} intensity images of the AR, taken at $\sim$16:00 UT, are presented in panels (a) and (b), respectively. In panel (c), the temporal variations in $\Delta H_{\rm R}$ (cross-solid line), the total unsigned flux (diamond-solid line), and GOES soft X-ray flux (dotted line) are shown, together with the two distinct (i.e., increase and steady) phases in the observed profile of $\Delta H_{\rm R}$. Figure reproduced from \citet{2008ApJ...686.1397P} by permission of the AAS.
\label{fig:park2008}}
\end{figure}

With respect to flare-associated variations in $F_{\rm tot}$ on a long-term scale, one of the most remarkable observational findings is that flaring ARs show a large, continuous increase in $\Delta H_{\rm R}$ (i.e., the time integral of $F_{\rm tot}$) over an interval of several hours to a few days before the flare onset. Figure~\ref{fig:park2008} presents an example of this long-term increasing trend of $\Delta H_{\rm R}$ (cross-solid line in panel c) in NOAA AR 9672, which produced an X1.3 flare (dotted line in panel c) on 2001 October 25. Interestingly, $\Delta H_{\rm R}$ increased monotonically during a pre-flare phase of $\sim$24 hour (marked by the dark gray bar and labeled I) and then remained almost constant for the post-flare phase (labeled II). On the other hand, AR's total unsigned magnetic flux (diamond-solid line in panel c) changed little over the entire investigation period, making it difficult to find any flare-associated temporal variations. This trend of increase in $\Delta H_{\rm R}$ was found in many other large flaring ARs \citep[e.g.,][]{2008ApJ...686.1397P,2010ApJ...720.1102P,2010ApJ...718...43P,2012ApJ...750...48P,2013ApJ...778...13P,2002ApJ...577..501K,2008PASJ...60.1181M,2012ApJ...761...86V,2014ApJ...794..118R,2015ApJ...806..245V,2015A&A...582A..55R}, whereas no characteristic evolution of $\Delta H_{\rm R}$ was observed in flare-quiet ARs. Some statistical studies of this increase in helicity in relation to the flaring activity are discussed in Section~\ref{subsec:applications_statistics}. 

\begin{figure}
\begin{center}
\includegraphics[width=0.95\textwidth]{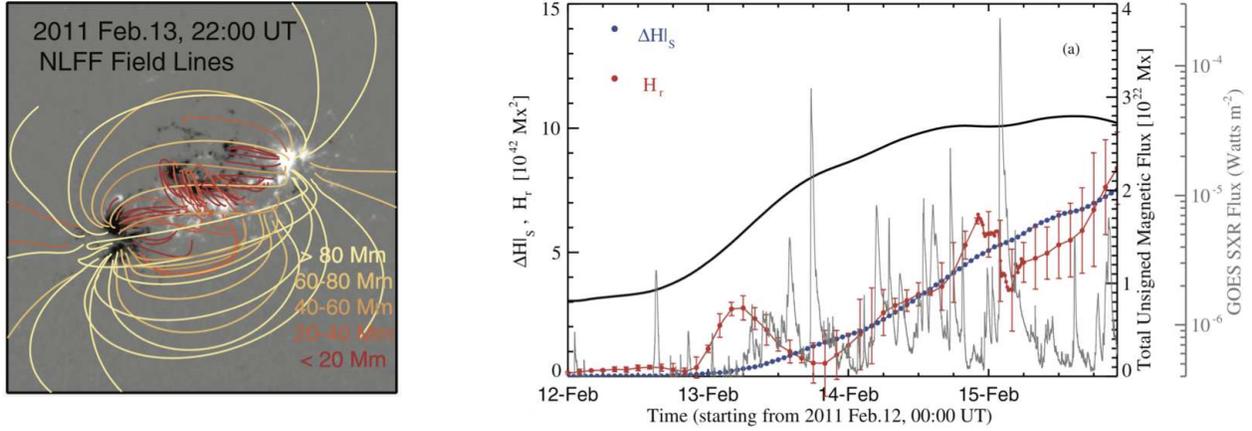}
\end{center}
\caption{Long-term increasing trend of $H_{\rm R}$ in the coronal volume of NOAA AR 11158 prior to an X2.2 flare on 2011 February 15. (Left) A selected set of coronal magnetic field lines extrapolated from the photospheric vertical magnetic field (grayscale image), is displayed with different colors according to their maximum altitude. (Right) Temporal variations of $H_{\rm R}$ (red line), $\Delta H_{\rm R}$ (blue line), the total unsigned magnetic flux (black line), and the GOES soft X-ray flux (gray line). Figure reproduced from \citet{2012ApJ...752L...9J} by permission of the AAS.
\label{fig:jing2012}}
\end{figure}

In the same context, as shown in Figure~\ref{fig:jing2012}, a long-term increase in helicity was also found in the time profile of the relative magnetic helicity $H_{\rm R}$ in the coronal volume of ARs that produced large flares \citep[e.g.,][]{2010ApJ...720.1102P,2012ApJ...752L...9J,2015RAA....15.1537J,2019ApJ...887...64T}. The observed increasing patterns of both $\Delta H_{\rm R}$ and $H_{\rm R}$ prior to flares support the idea that $H_{\rm R}$ in the AR corona is mainly supplied by $F_{\rm tot}$ through the bottom surface (i.e., the AR photosphere). It is also thought that $H_{\rm R}$ is well preserved over timescales of a few hours to several days, for example, as shown by \citet{2010ApJ...720.1102P}, based on the comparison of approximately five-day profiles between $H_{\rm R}$ and $\Delta H_{\rm R}$ for NOAA AR 10930. Moreover, over the course of AR evolution, the pre-flare increasing trend of the helicity may reflect a complicated process of building up the free magnetic energy in the corona by the emergence of twisted magnetic flux tubes and/or footpoint shear motions of the magnetic field line.

\begin{figure}
\begin{center}
\includegraphics[width=\textwidth]{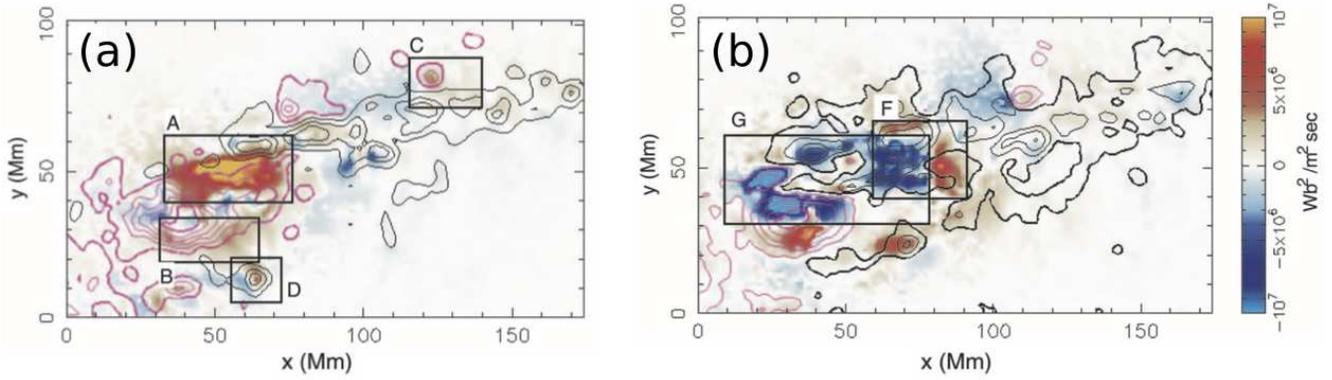}
\end{center}
\caption{Maps of the relative magnetic helicity flux density for NOAA AR 8100, observed at 00:49 UT (panel a) and 23:11 UT (panel b) on 1997 November 3, are presented in the blue and red color scale. The AR produced an X2.1 flare at 05:52 UT on November 4. The positive and negative vertical magnetic fields are indicated by pink and black contours, respectively. Some regions of interest are marked by the rectangular boxes with the following labels. A: the main PIL; B: a southern part of the main positive polarity sunspot; C and D: satellite patches in which the magnetic polarity is opposite to that of the nearby main spot; F: a region with strong upflows on the photospheric surface; G: a region in which the helicity flux density is predominantly negative in sign. Figure reproduced from \citet{2002ApJ...577..501K} by permission of the AAS.
\label{fig:kusano2002}}
\end{figure}

\begin{figure}[t!]
\begin{center}
\includegraphics[width=0.8\textwidth]{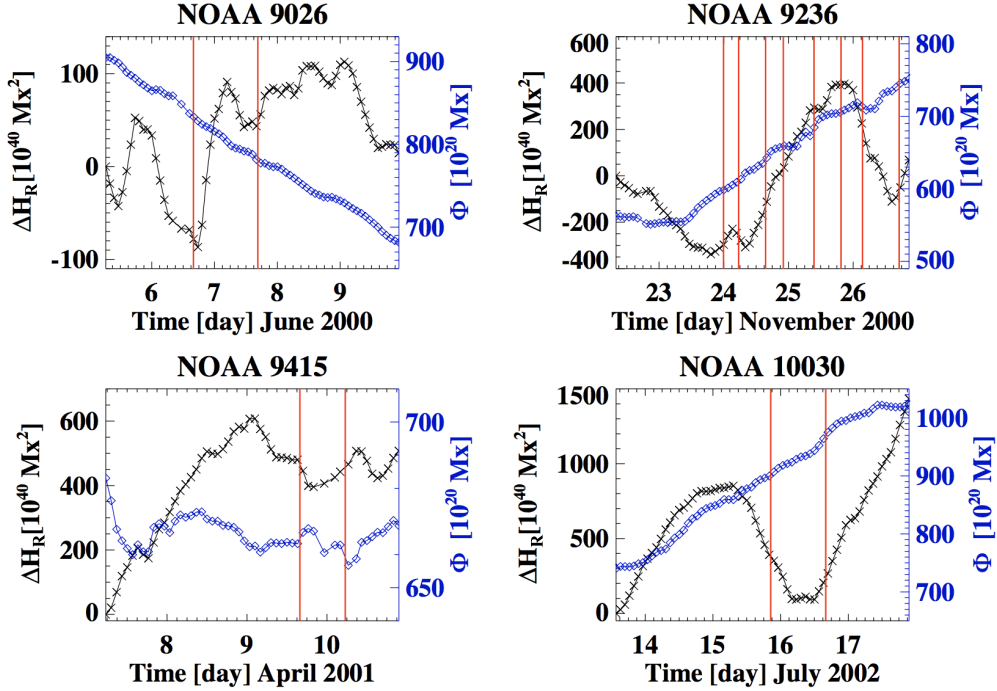}
\end{center}
\caption{Time profiles of $\Delta H_{\rm R}$ (black line) and $\Phi$ (blue line) for four ARs that produced multiple eruptive flares with CMEs. The red vertical lines indicate the times when the CMEs first appeared in the LASCO/C2 field of view. Figure reproduced from \citet{2012ApJ...750...48P} by permission of the AAS.
\label{fig:park2012}}
\end{figure}

One of the most important aspects of studying the magnetic helicity is that it measures the handedness of twisted magnetic field lines: i.e., {\it positive} helicity in the case of right-handed twist and {\it negative} helicity for the left-handed case. As shown in Figure~\ref{fig:kusano2002}, \citet{2002ApJ...577..501K} examined the spatial distribution and evolution of the relative magnetic helicity flux density across the entire photospheric surface of NOAA AR 8100 during a five-day interval from 1997 November 1 to 5. They found that negative helicity transport was predominantly caused by $F_{\rm v_{n}}$ over the first 2.5 days of the investigated interval, while helicities of positive and negative signs were transported by $F_{\rm v_{n}}$ and $F_{\rm B_{n}}$, respectively, for the rest of the interval. In the latter phase, multiple flares, consisting of three M-class flares and one X2.1 flare, occurred in the region (labeled G in panel b of Figure~\ref{fig:kusano2002}), where negative helicity was consistently transported primarily owing to the shear motion along the main PIL. Such a characteristic follow-up transport of helicity via a localized area (typically around PILs) with a sign opposite to the preloaded helicity in the AR corona has been reported in many other observations of flaring ARs \citep[e.g.,][]{2003AdSpR..32.1949Y,2004ApJ...615.1021W,2010ApJ...720.1102P,2012ApJ...761...86V,2015RAA....15.1537J}. When the subsequent opposite-sign helicity flux in the localized areas of the AR becomes larger than the helicity flux in the rest of the areas, the sign of $F_{\rm tot}$ is actually reversed over the course of its measurement, which is called the {\it helicity sign reversal}.

\citet{2012ApJ...750...48P} performed time series analysis of $F_{\rm tot}$ for a set of different AR samples, each of which produced eruptive flares and accompanying CMEs. They found that all the investigated eruptive flares occurred following a remarkable increase in $\Delta H_{\rm R}$ first and the subsequent reversal of the helicity sign (see Figure~\ref{fig:park2012}). Here the helicity sign reversal is shown as the slope of the time profile of $\Delta H_{\rm R}$ changes its sign. In contrast, examining temporal variations of $F_{\rm tot}$ for NOAA AR 12257, \citet{2021MNRAS.507.6037V} reported that there were only small C-class flares without CMEs even when the helicity sign reversal proceeded. However, it should be noted that in the case of NOAA AR 12257, there was no significant transport of helicity before the sign reversal, while a large amount of helicity transport was made for the ARs under study in \citet{2012ApJ...750...48P}. A careful investigation is needed when interpreting the helicity flux density map; for example, artificial signals may be generated owing to various limitations in observational data and practical methods, as indicated by \citet{2005A&A...439.1191P,2006A&A...452..623P}. Meanwhile, \citet{2015RAA....15.1537J} conducted an interesting comparison study of eruptive versus confined flares by analyzing the evolution of $H_{\rm R}$ in the AR corona. They found that $H_{\rm R}$ remarkably decreased $\sim$4 hour prior to the eruptive X2.2 flare on 2011 February 15, whereas no such decrease in $H_{\rm R}$ was found in the case of the confined X3.1 flare on 2014 October 24.

It is noteworthy that the observed signatures of (1) the helicity sign reversal and (2) the sudden decrease in $H_{\rm R}$ prior to eruptive flares can be explained by the flare-trigger MHD simulation model of \citet{2012ApJ...760...31K}, in which, as a flare trigger, a small-scale bipolar magnetic structure supplies helicity of opposite sign into the pre-existing sheared arcade system.

\begin{figure}[t!]
\begin{center}
\includegraphics[width=0.95\textwidth]{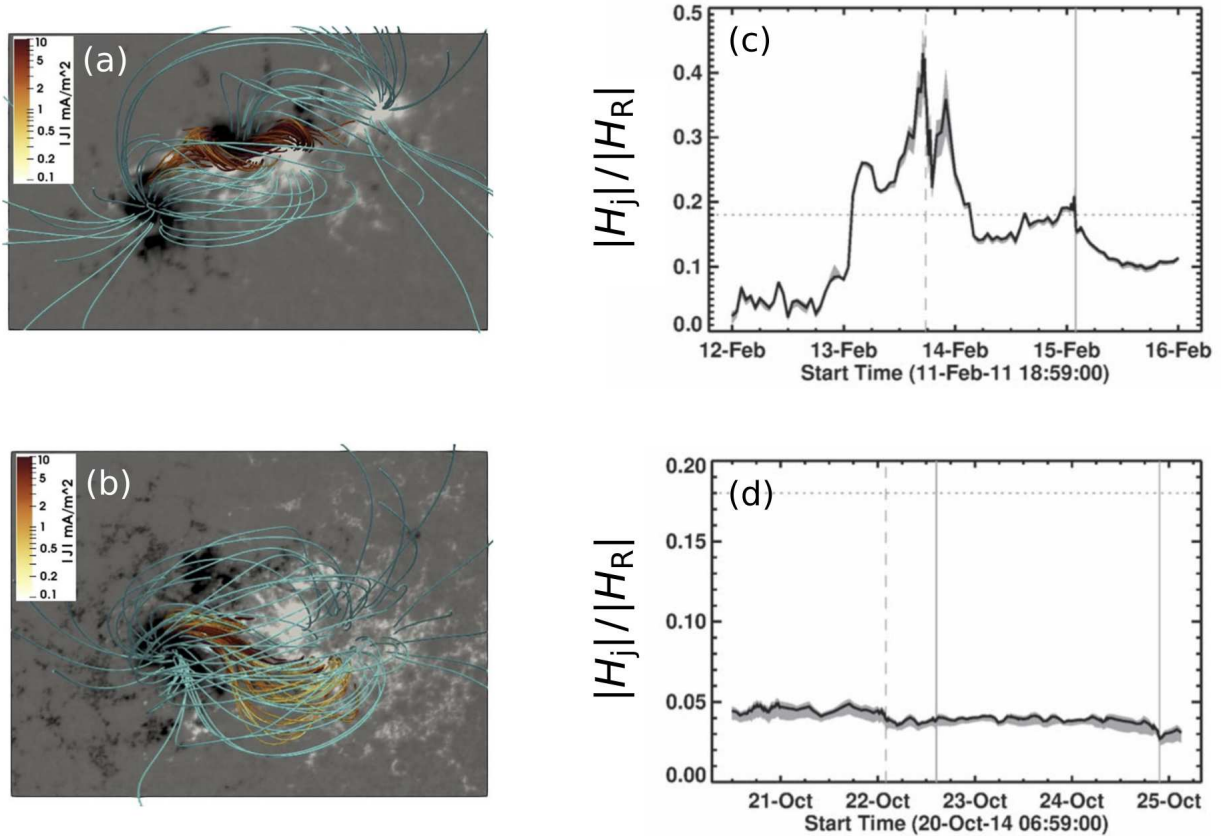}
\end{center}
\caption{(Left) Coronal magnetic field lines extrapolated from the vertical magnetic field (grayscale image) at the photospheric surface of (a) AR 11158 at 21:00 UT on 2011 February 14 and (b) AR 12192 at 19:00 UT on 2014 October 24. (Right) Time profiles of $|H_{\rm j}|/|H_{\rm R}|$ for (c) AR 11158 and (d) AR 12192. In panels (c) and (d), the shaded areas represent the spread of the values of $|H_{\rm j}|/|H_{\rm R}|$ from the three different methods used to estimate the volume helicity. The vertical dashed and solid lines indicate the GOES peak times of the M- and X-class flares, respectively. The horizontal dashed line marks the value of $|H_{\rm j}|/|H_{\rm R}|=0.17$ as a reference. Figure reproduced from \citet{2019ApJ...887...64T} by permission of the AAS.
\label{fig:thalmann2019}}
\end{figure}

There have been longstanding discussions on the role of magnetic helicity as a sufficient indicator of solar eruptions. Based on practical methods of estimating the magnetic helicity (refer to Section~\ref{sec:methods}), many trials have been conducted so far to determine the {\it magic} parameter that has the capability of providing a threshold above which flares and CMEs occur. In this respect, \citet{2017A&A...601A.125P} proposed a new parameter for the solar eruptivity criterion, which is the ratio of the magnetic helicity of the current-carrying magnetic field to the total relative helicity, $|H_{\rm j}|/|H_{\rm R}|$. For NOAA AR 12673, which produced the largest X-class flare of solar cycle 24, \citet{2019A&A...628A..50M} found that the helicity ratio increased and then relaxed to lower values before and after, respectively, each of the two major X2.2 and X9.3 flares on 2017 September 6. Through the analysis of observations and coronal field extrapolations of two flaring ARs with eruptive versus confined flares, as shown in Figure~\ref{fig:thalmann2019}, \citet{2019ApJ...887...64T} found a similar increasing/decreasing trend of $|H_{\rm j}|/|H_{\rm R}|$ before and after the major eruptive flares in NOAA AR 11158. However, this was not observed in NOAA AR 12192, which produced only confined flares without CMEs. Another follow-up study of \citet{2021A&A...653A..69G}, with a set of 10 different NOAA-numbered AR samples, shows results supporting that the helicity ratio has a strong ability to indicate the eruptive potential of an AR, albeit with a few exceptions. A further statistical study with more samples is needed.

Then, why does the helicity ratio, $|H_{\rm j}|/|H_{\rm R}|$, seem to predict flare eruptions well? There have been intensive studies on critical conditions for a current-carrying structure (e.g., twisted flux ropes) to successfully erupt against its surrounding confinement field, as represented by the torus instability \citep{2006PhRvL..96y5002K,2010ApJ...718.1388D} and the double-arc instability \citep{2017ApJ...843..101I}. In fact, various ``relative'' quantities have been proposed in the framework of instabilities for a current system in the corona and tested with observations of the eruptive and confined flares \citep{2015ApJ...804L..28S,2017ApJ...834...56T,2020ApJ...894...20L,2020ApJ...900..128L,2020Sci...369..587K,2022ApJ...926...56K}. In this respect, $|H_{\rm j}|/|H_{\rm R}|$ may reflect the magnetic relationship between the current-carrying structure (the flux rope) and the entire system (the whole AR), which is one of the key factors that determine these instabilities.

Recently, using wavelet analysis, attempts have been made to determine any characteristic periodicities in the time series of $F_{\rm tot}$ for flaring ARs. For example, \citet{2020ApJ...897L..23K} and \citet{2021arXiv211205933S} reported that there are some differences in the identified periodicities between flaring ARs with large M- and X-class flares and ARs with smaller B- and C-class flares. In this context, more detailed studies in the future may help us to better understand whether the evolution of the magnetic field on the photosphere is inherently different between large flaring ARs and the others, as well as to predict the likelihood of flare occurrence for a target AR.

\subsection{Statistical trends}\label{subsec:applications_statistics}

Examination of the magnetic helicity on many sample ARs has revealed the statistical tendency that flare-rich ARs harbor a larger amount of magnetic helicity or helicity flux. For instance, \citet{2007ApJ...671..955L} surveyed the helicity injections for 48 X-flaring ARs and 345 reference regions without X-flares. They derived an empirical threshold for the occurrence of an X-class flare that the peak helicity flux, $\max{\lvert\langle F_{\rm tot}\rangle\rvert}$ given by Equation (\ref{eq:chae_helicityflux}), should exceed a magnitude of $6\times 10^{36}\ {\rm Mx}^2\ {\rm s}^{-1}$.

\begin{figure}[t!]
\begin{center}
\includegraphics[width=0.75\textwidth]{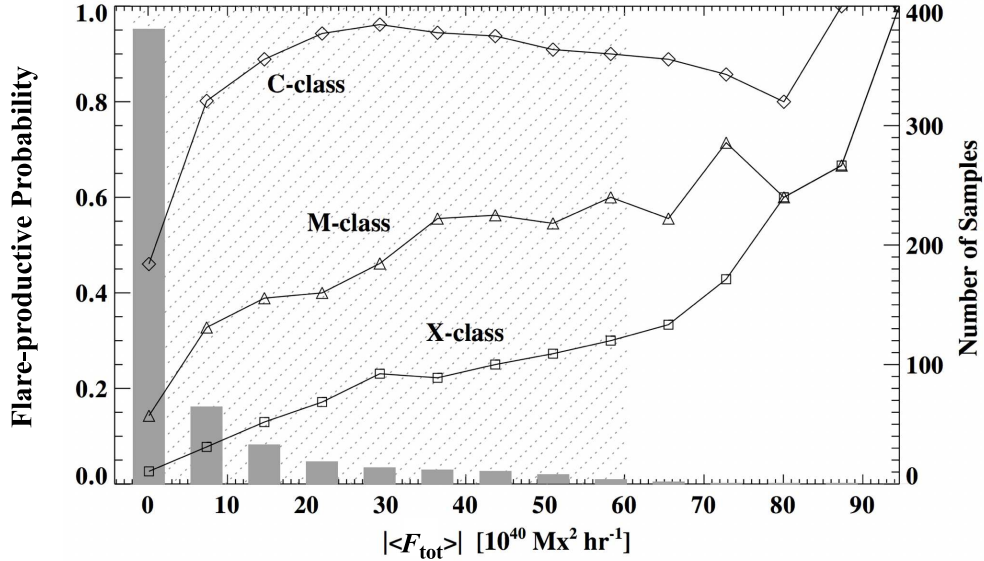}
\end{center}
\caption{Flare-productive probability as a function of $\lvert\langle F_{\rm tot}\rangle\rvert$, respectively, for C-class (diamonds), M-class (triangles), and X-class (squares). The gray bars indicate the total number of AR samples in the defined bins. Figure reproduced from \citet{2010ApJ...718...43P} by permission of the AAS.
\label{fig:park2010}}
\end{figure}

\citet{2010ApJ...718...43P} expanded the analysis to 378 ARs and investigated the correspondence between the 24 hour-averaged helicity flux, $\lvert\langle F_{\rm tot}\rangle\rvert$ based on Equation (\ref{eq:chae_helicityflux}), and the next 24-hour flare index (i.e., the sum of GOES soft X-ray peak intensities of flares that occurred in the next 24 hour after the helicity measurement). They demonstrated that, for 91 subsamples with an unsigned magnetic flux of $(3$--$5)\times 10^{22}\ {\rm Mx}$, the helicity flux for the flaring ARs was approximately twice that of the quiescent ARs. On the other hand, 118 ARs with large helicity rates did not show a significant difference in magnetic flux between the flaring and quiescent groups. As shown in Figure~\ref{fig:park2010}, these authors also found that the larger value of $\lvert\langle F_{\rm tot}\rangle\rvert$ an AR has, the higher the flaring activity of the AR is in all cases of C-, M-, and X-class flares. 

Regarding the differences between CME-eruptive and non-eruptive flares, \citet{2004ApJ...616L.175N} examined the coronal magnetic helicity, derived under the assumption that the coronal field of each AR has a constant force-free $\alpha$ (Section \ref{subsec:extrapolation}) for 133 M-class events from 78 ARs. They revealed the statistical tendency that the pre-flare $\alpha$ values and coronal helicity of the non-eruptive flares are smaller than those of the eruptive flares. \citet{2012ApJ...750...48P} examined 47 CMEs in 28 ARs and showed that there was also good correlation between $\lvert\langle F_{\rm tot}\rangle\rvert$ and the CME speed. Interestingly, by tracing the helicity evolution, the CMEs analyzed were divided into two groups: a group in which helicity increased monotonically with one sign of helicity, and a group in which the sign of helicity reverses after a significant helicity injection (reversed shear: \citealt{2004ApJ...610..537K,2012ApJ...760...31K}).

\begin{figure}[t!]
\begin{center}
\includegraphics[width=0.7\textwidth]{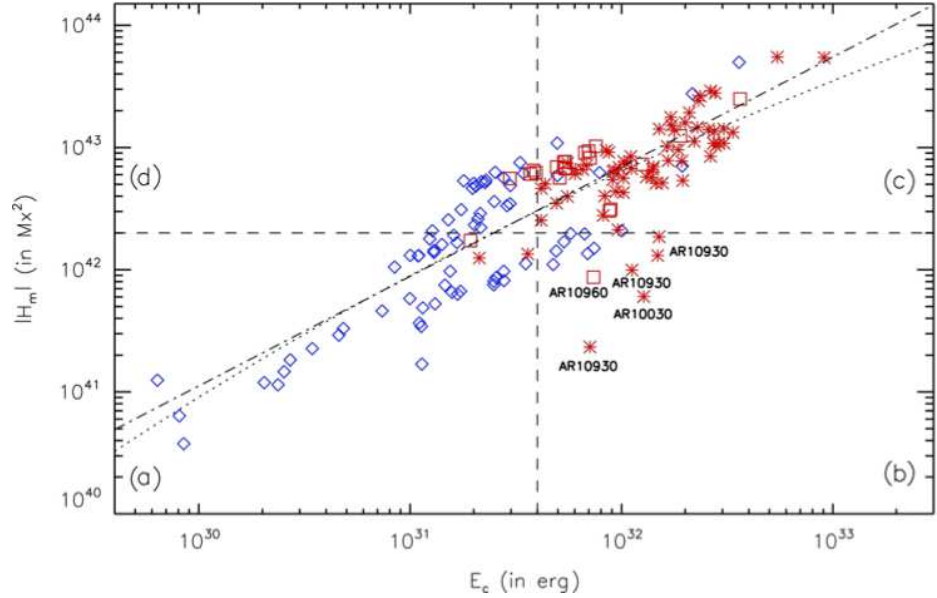}
\end{center}
\caption{Scatter plots of ARs as functions of free magnetic energy (denoted as $E_{\rm c}$ in this diagram) and absolute relative magnetic helicity ($|H_{\rm m}|$ here). The blue diamonds, red squares, and red asterisks denote the non-flaring, M-class flaring, and X-class flaring ARs, respectively. The dashed horizontal and vertical lines indicate the thresholds for M-class events ($2\times 10^{42}\ {\rm Mx}^{2}$ and $4\times 10^{31}\ {\rm erg}$). The dotted and dash-dotted lines are the least-square best fit and least-square best logarithmic fit for the data points, respectively. Figure reproduced from \citet{2012ApJ...759L...4T} by permission of the AAS.
\label{fig:tziotziou2012}}
\end{figure}

\citet{2012ApJ...759L...4T} investigated the relationship between the free magnetic energy ($\Delta E_{\rm mag}$) and relative helicity ($H_{\rm R}$) on 42 ARs, finding that the flaring ARs possess higher free energy and relative helicity. As demonstrated in Figure \ref{fig:tziotziou2012}, there exists a monotonic scaling between the two variables with a threshold for M-class events, which is $4\times 10^{31}\ {\rm erg}$ and $2\times 10^{42}\ {\rm Mx}^{2}$ for the energy and helicity, respectively. This energy-helicity diagram was further examined by \citet{2014A&A...570L...1T}. By using 3D numerical simulations of eruptive and non-eruptive cases and two observed, eruptive and non-eruptive ARs, they reconstructed the energy-helicity diagram and found a consistent monotonic scaling law.

However, care must be taken when considering such statistical results because magnetic helicity is homogeneous to the square of the magnetic flux. As shown in Section \ref{sec:intro}, ARs with larger areas or magnetic fluxes are more prone to flares, which may be included in the statistical results using extensive parameters (those scaling with AR size) including magnetic helicity \citep{2009ApJ...705..821W,2015ApJ...798..135B,2015ApJ...804L..28S,2017ApJ...850...39T}. Therefore, to assess the eruptivity of ARs purely owing to the structural complexity and not to the AR area, it is worth considering the helicity normalized by the flux squared, $H_{\rm R}/\Phi^{2}$ \citep{2009AdSpR..43.1013D}. On the other hand, the results of statistical studies clearly demonstrate that there exist quantitative differences between flare-productive and quiescent ARs. This suggests that helicity-related variables may be useful for predicting and forecasting imminent flare occurrences.

\subsection{Magnetic tongues: A proxy for the helicity sign}\label{subsec:applications_tongues}

It has been suggested that {\it magnetic tongues} in ARs, the elongated yin-yang shaped patches of positive and negative magnetic polarities on the solar surface, can be used as a proxy for the magnetic helicity \citep{2000ApJ...544..540L}. It is thought that the positive and negative polarities extended on both sides of the PIL in longitudinal (or line-of-sight) magnetograms reflect the vertical projection of the poloidal (azimuthal) component of the emerging magnetic flux tube. When an arched flux tube rises and penetrates the photospheric surface, the horizontal cross section of the tube displays a pair of opposite polarity patches (magnetic tongues) with the PIL in between. If the flux tube is twisted or writhed, that is, if the emerging flux has a non-zero magnetic helicity (the situation in panels (a1) or (a4) in Figure \ref{fig:dhdt_diagram}), the tongue structure has an axisymmetric pattern, and the PIL deviates from an orthogonal angle to the main bipolar axis. Therefore, the layout of the tongues and the direction of the PIL were used to determine the sign of the magnetic helicity of the ARs. In fact, flux emergence simulations that assume twisted flux tubes often display magnetic tongues in the photosphere \citep[e.g.,][]{2010A&A...514A..56A,2011PASJ...63..407T}.

\citet{2009SoPh..258...53C} analyzed the evolution of the magnetic field in NOAA AR 10365 in May 2003 and found that it had a positive magnetic helicity. In fact, this AR produced an M1.6-class flare leaving a pair of J-shaped flare ribbons, another indication of twisting in ARs. \citet{2014SoPh..289.2041M} analyzed the force-free parameter, $\alpha$ (Section \ref{subsec:extrapolation}), for NOAA ARs 11121 and 11123 and confirmed that the positive value for $\alpha$ agrees with the observed magnetic tongue.

Statistical analysis of magnetic tongues by \citet{2011SoPh..270...45L} on 40 ARs showed that the helicity signs determined from the tongues were consistent with those derived from the photospheric helicity fluxes. \citet{2015SoPh..290..727P,2015SoPh..290.3279P} systematically surveyed 41 bipolar ARs and demonstrated that the twist estimated from the tongues in general matched the force-free $\alpha$ derived from the coronal field extrapolations. By analyzing 187 ARs, \citet{2016SoPh..291.1625P} further investigated the dependence of tongues on the activity cycle. They found that the angle of the PIL relative to the bipolar axis (AR tilt) has only a weak sign dominance in each solar hemisphere ($\gtrsim 50$\%). In addition, the characteristics of the tongues, not only the PIL angle but also the amount of magnetic flux, polarity size, latitude, emergence rate, etc., are not dependent on the periods of the cycle.

\section{Numerical models}\label{sec:simulations}

\subsection{Coronal field reconstructions}\label{subsec:extrapolation}

\begin{figure}
\begin{center}
\includegraphics[width=0.8\textwidth]{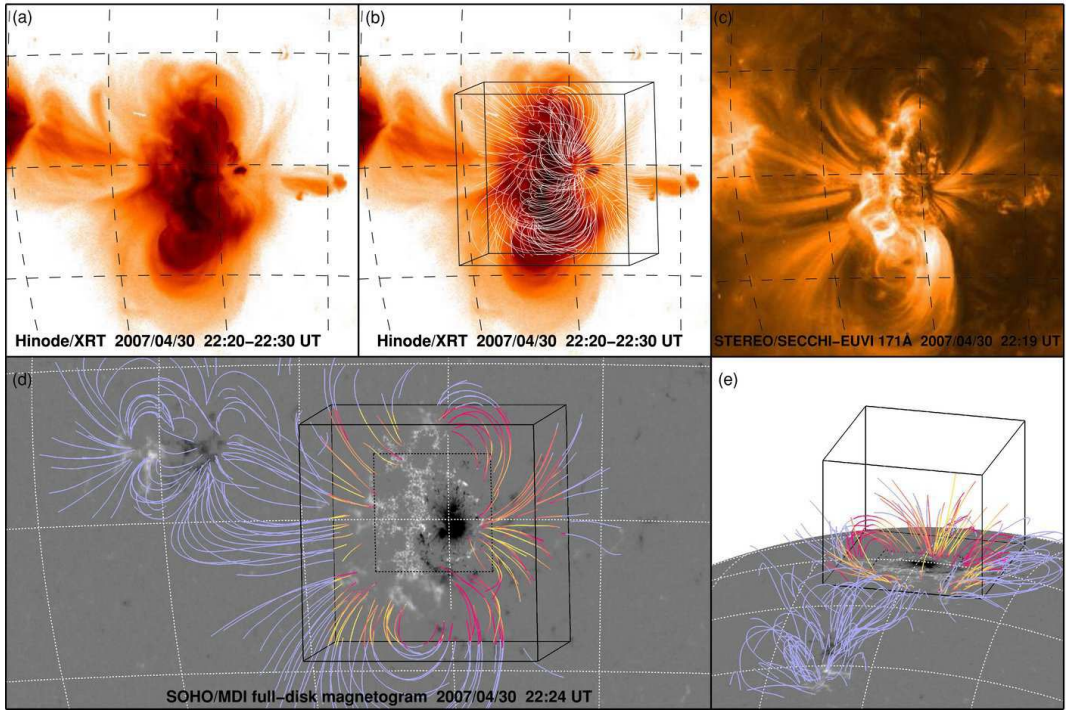}
\end{center}
\caption{NLFFF extrapolation on NOAA AR 10953. The best-fit NLFFF result is shown in panel (b) as white lines. The stereoscopically reconstructed field lines are displayed in panels (d) and (e), in which the solid black cubes outline the NLFFF computational domain. The loops inside the box are colored according to their misalignment angle from the NLFFF solution (yellow to orange corresponds to 5$^{\circ}$ to 45$^{\circ}$). Figure reproduced from \citet{2009ApJ...696.1780D} by permission of the AAS.
\label{fig:derosa2009}}
\end{figure}

Owing to the limitations of magnetic field measurements in the solar corona, we rely on coronal field models to estimate the magnetic field energy and helicity. The simplest approach is to extrapolate the coronal field based on the observed photospheric magnetic field under the assumption that non-magnetic forces (e.g., pressure gradient and gravity) are negligible and that the Lorentz force vanishes, i.e.,
\begin{equation}
\boldsymbol{j}\times\boldsymbol{B}=0,
\end{equation}
where $\boldsymbol{j}=c/(4\pi)(\nabla\times\boldsymbol{B})$ is the electric current. This force-free condition can also be written as
\begin{equation}
\nabla\times\boldsymbol{B}=\alpha\boldsymbol{B},
\label{eq:forcefreecondition}
\end{equation}
where $\alpha$ is the force-free parameter. If $\alpha=0$, or equivalently $\boldsymbol{j}=0$ (current-free), the coronal field is the potential field ($\boldsymbol{B}_{\rm p}$), which is often used as the reference field when estimating the relative magnetic helicity (as in Equation (\ref{eq:relativehelicity})). If $\alpha$ is non-zero and constant everywhere in the coronal volume under consideration, the magnetic field is called a linear force-free field (LFFF). If $\alpha$ is non-uniform, the field is a non-linear force-free field (NLFFF: Figures \ref{fig:jing2012}, \ref{fig:thalmann2019}, and \ref{fig:derosa2009}).

Here, only the vertical component of the photospheric magnetic field ($B_{\rm n}$) is required as the bottom boundary condition to obtain the potential field, whereas all components of the vector magnetogram ($\boldsymbol{B}$) are used for the LFFF and NLFFF. Extrapolation models, particularly NLFFF methods, have been extensively used for the direct determination of volume helicity in the AR corona (see discussions in the previous sections). The readers are referred to, e.g., \citet{2016PEPS....3...19I} and \citet{2021LRSP...18....1W} for detailed accounts of extrapolation methods.

These reconstruction models, however, only provide steady-state coronal fields and may not be applicable to flare-productive ARs, which are highly dynamic in nature. To overcome this issue, temporally evolving models have been used. One approach of such coronal field reconstructions is data-constrained models, where the initial coronal field is prepared by extrapolations based on a snapshot magnetogram, and the subsequent dynamical evolution is obtained by solving the MHD equations.

\begin{figure}
\begin{center}
\includegraphics[width=0.9\textwidth]{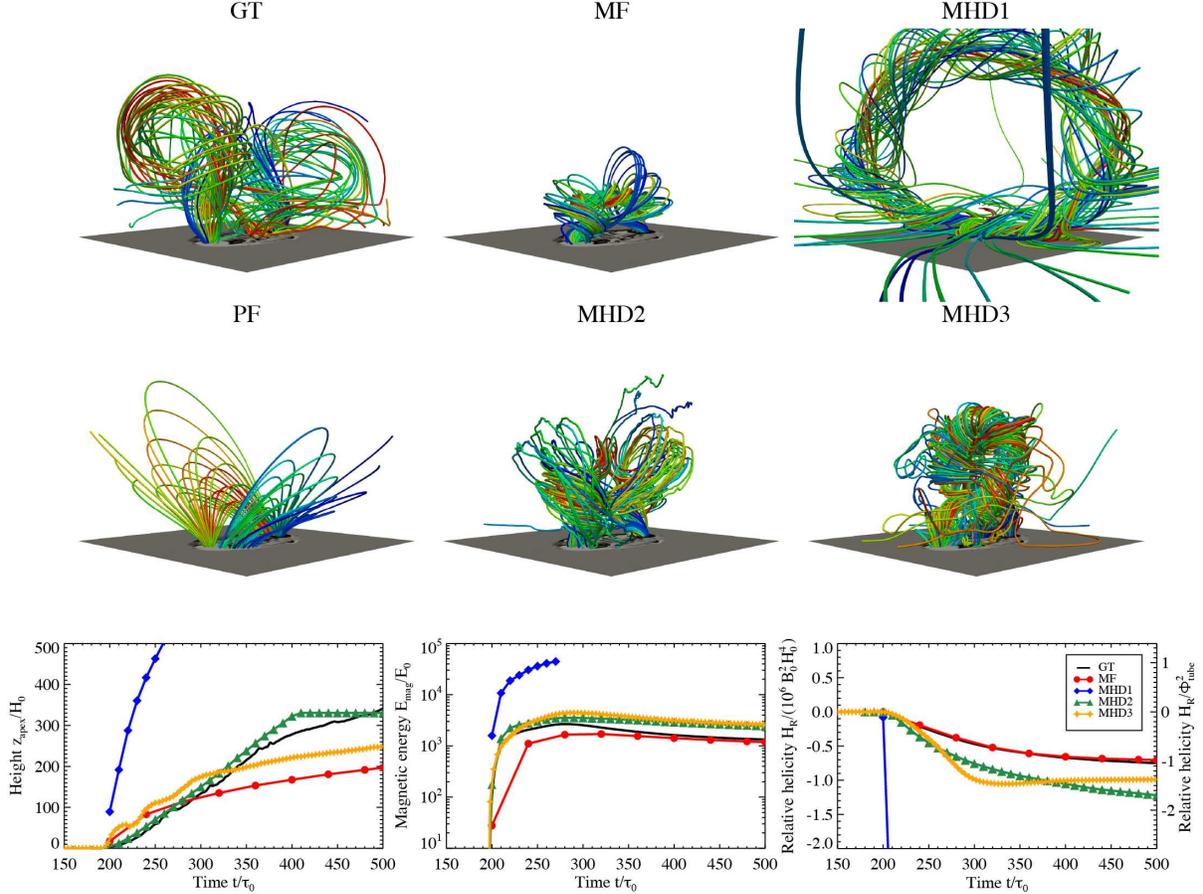}
\end{center}
\caption{(Top and middle) Magnetic fields for ground-truth flux-emergence simulation (GT), potential field extrapolation (PF), and four data-driven coronal field reconstructions (MF, MHD1, MHD2, and MHD3). The bottom boundary shows the vertical GT magnetic field (i.e., the photospheric magnetogram). The colored tubes indicate the field lines, where the tubes with reddish (bluish) colors are integrated from the seeds placed in the positive (negative) polarity. The seeds were identical in all six cases. The time for MHD1 is 1000 s after the photospheric appearance of the emerging flux, while all other times are 4000 s. (Bottom) Temporal evolution of the apex height $z_{\rm apex}$, magnetic energy $E_{\rm mag}$, and relative magnetic helicity $H_{\rm R}$ for GT and the four coronal field models, MF, MHD1, MHD2, and MHD3. The normalization factor for time $\tau_{0}$ corresponds to 25 s. Figure reproduced from \citet{2020ApJ...890..103T} by permission of the AAS.
\label{fig:toriumi2020}}
\end{figure}

Data-driven models, in which the coronal field evolves in response to sequentially updated bottom boundary magnetograms, may provide even more realistic coronal reconstructions. To investigate how accurately different data-driven models can reproduce a coronal field and quantities such as the magnetic helicity, \citet{2020ApJ...890..103T} used a flux emergence simulation as the ground truth (GT) dataset \citep{2017ApJ...850...39T}. In this GT simulation, a twisted flux tube initially placed in the convection zone rises into the atmosphere and eventually builds up coronal loops. These authors compared the GT coronal field with those reconstructed using multiple data-driven models based on sequential GT magnetograms. As illustrated in Figure \ref{fig:toriumi2020}, the helical flux rope structure was reproduced in all coronal field models. On the other hand, model-dependent results were obtained for quantitative comparison, with the magnetic energies ($E_{\rm mag}$ and $\Delta E_{\rm mag}$) and relative magnetic helicity ($H_{\rm R}$) varying from the cases almost comparable to GT to those differed by orders of magnitude. The magnetic helicity was derived by the method in \citet{2000ApJ...539..944D}, following \citet{2012SoPh..278..347V}. The observed model discrepancies were attributed to a highly non-force-free input bottom boundary (GT magnetograms) and to modeling treatments of the background atmosphere, bottom boundary, and spatial resolution.

\subsection{Idealized simulations}

Numerical models of flux rope eruptions (i.e., CMEs) have also been employed to test the concept of magnetic helicity. For instance, \citet{2009ApJ...702..580M} used eruption models to show that the additive self helicity ($H_{{\rm R}i}$: Equation (\ref{eq:additiveselfhelicity1})) is useful for determining the threshold beyond which a flux rope becomes unstable. The additive self helicity divided by the square of the magnetic flux through a cross section of the tube, $H_{{\rm R}i}/\Phi^2$, can be considered as a generalization of the twist number \citep{1984JFM...147..133B}, which has been used to discuss the threshold of kink instability. The authors investigated the evolution of $H_{{\rm R}i}/\Phi^2$ in the simulations, in which a twisted flux rope kinematically inserted into the computational domain interacted with a pre-existing coronal arcade and deformed into a writhed configuration owing to the kink instability \citep{2003ApJ...589L.105F}. Over time, the (absolute) magnitude of $H_{{\rm R}i}/\Phi^2$ increased until it reached a threshold value of $\sim$1.5 for the instability to occur.

\begin{figure}
\begin{center}
\includegraphics[width=0.75\textwidth]{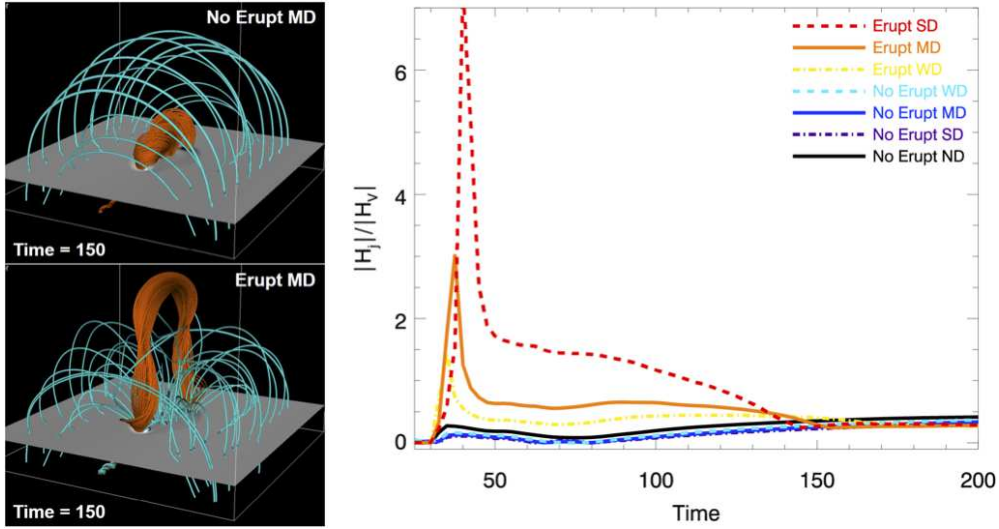}
\end{center}
\caption{Examination of helicity ratio ($|H_{\rm j}|/|H_{\rm R}|$) by \citet{2017A&A...601A.125P}. (Left) Snapshots of two simulation cases (non-eruptive and eruptive) with medium arcade strength. (Right) Time evolution of the helicity ratio for the seven parameter cases. The eruptive cases show higher values above 0.45. Note that the relative helicity $H_{\rm R}$ is denoted in the figure as $H_{\rm V}$. Reproduced with permission from Astronomy \& Astrophysics, \textcopyright ESO.
\label{fig:pariat2017}}
\end{figure}

Another simulation model that was employed to validate the usage of helicity was given by \citet{2013ApJ...778...99L,2014ApJ...787...46L}, where a magnetic flux emerges into a magnetized corona, which may or may not erupt depending on the direction of the coronal arcade (Figure \ref{fig:pariat2017}). \citet{2017A&A...601A.125P} investigated the temporal evolution of several global scalar quantities to determine whether they could discriminate between eruptive and non-eruptive cases. It was found that the helicity ratio $|H_{\rm j}|/|H_{\rm R}|$ can determine the eruptivity, whereas other quantities such as the magnetic flux ($\Phi$), magnetic energies ($\Delta E_{\rm mag}$, $\Delta E_{\rm mag}/E_{\rm mag}$, etc.), and total relative helicity ($|H_{\rm R}|$) cannot (see Section \ref{subsec:applications_tempvar} for the details).

\section{Summary and Discussion}\label{sec:discussion}

In this chapter, we have reviewed the following: (1) solar flares and CMEs tend to occur in ARs with complex structures; (2) therefore, various concepts of magnetic helicity, which is the physical quantity that measures the complexity of magnetic fields, help to understand the occurrence of flares and CMEs; and (3) based on the statistical findings, these concepts may be useful for the prediction and forecasting of such eruptive events. Magnetic helicity is not limited to mere theoretical research, but its practicality has been proven by a variety of studies that exploit actual observational data. However, there remain important topics that cannot be addressed in detail in this chapter.

For instance, there is an issue that the absolute value of magnetic helicity may vary, depending on which gauge is employed for both $\boldsymbol{A}$ and $\boldsymbol{A}_{\rm p}$, as discussed in \citet{2015A&A...580A.128P} and \citet{2019ApJ...882..151S}. This gauge issue can be problematic, particularly when comparing results obtained from different methods of estimating helicity and investigating a critical value of helicity for flare- or CME-trigger instabilities. In addition, as in Equation (\ref{eq:dhdt_original_decomp}), the definition of the temporal variation of the relative magnetic helicity $dH_{\rm R}/dt$ and its flux components on the boundary surface may not be fully consistent between previous studies with somewhat different gauge choices. We also note that each flux component may not be gauge-invariant, but only their sum is. On the other hand, different ways of implementing helicity formulas, as well as in preprocessing observational data, can result in inconsistent results due to numerical errors and artificial noise. Future community-wide efforts, such as ``Helicity Thinkshops'' and ``Helicity2020'', may need to be made to discuss and achieve a general consensus on the issues addressed here.

It is important to note that we still need to resolve various sources of uncertainty in determining the evolution of magnetic helicity for ARs. First, as a problem on the observational side, there are inaccuracies in photospheric magnetograms. For instance, if there are magnetic field structures with the sizes smaller than the spatial resolution, magnetic helicity may be underestimated. Other causes of inaccuracies include the 180 degree ambiguity of the transverse field, general difficulties in calibration, improperly assumed atmospheres in the Stokes inversions, corrugation of line formation heights, and inadequate sampling of spectral profiles (see, e.g., \citealt{2003isp..book.....D}; some of these issues will be discussed later in this section). Second, there is a difficulty of using the photospheric data in estimating the relative helicity and/or helicity flux for the corona. The dynamic chromosphere lying between the photosphere and the corona may have an effect on the coronal energetics that cannot be perceived by photosphere observation alone, and this may also be one of the causes of misassessing magnetic helicity of the corona. Third, as mentioned in the previous paragraph, because the relative helicity is not invariant to gauge selection, it is critically important to confirm that the applied gauges are the same when comparing different relative helicity estimates. Otherwise, the comparison will lead to a wrong conclusion about the AR evolution and its relationship with eruptivity (see Sections \ref{sec:concepts} and \ref{subsec:methods_volumehelicity} for the details).

To understand the flaring activity of ARs, thus far, most magnetic helicity studies have focused on the temporal evolution of magnetic helicity with a set of {\it individual} ARs over a span (typically, a few days to two weeks) of their (partial) passages on the solar disk. On the other hand, very recently, a novel approach was designed by \citet{2021ApJ...911...79P} to examine whether there are any flare-associated differences in heliographic regions on a larger scale with respect to the degree of complying with the hemispheric sign preference (HSP) of magnetic helicity. The HSP of magnetic helicity has a unique cycle-independent tendency, in which various features in the northern hemisphere of the Sun tend to exhibit negative or left-handed helicity, while they are positive in the southern hemisphere. In the study of \citet{2021ApJ...911...79P}, heliographic regions were defined in the Carrington longitude-latitude plane with a grid spacing of 45$^{\circ}$ in longitude and 15$^{\circ}$ in latitude. The authors examined the relative magnetic helicity flux $F_{\rm tot}$ across the photospheric surface for 4,802 samples of 1,105 unique ARs observed during solar cycle 24. They found that heliographic regions with lower degrees of HSP compliance are likely to show higher levels of flaring activity as well as larger values of the average magnetic flux. What does this observational result tell us? Based on simulations of the rise of magnetic flux tubes deep in the convection zone by \citet{2021ApJ...909...72M}, we can deduce that if the helicity sign of a rising flux tube is against the HSP, then the tube's field strength is relatively larger for its successful rise compared to the other case of a tube with its helicity sign following the HSP. The assumption behind this speculation is that the background poloidal field strength is quasi-isotropic in the tachocline. The association of the lower HSP with the higher flaring activity may also suggest that there are localized regions in the convection zone where large-scale downflows and strong turbulence are present. Rising flux tubes therein may have an opposite sign of magnetic helicity as compared to the one expected from the HSP, may possess greater magnetic complexity (such as $\delta$-spots), and may produce large flares. Such HSP studies may offer an indirect understanding of the elusive magnetic nature of the solar interior as well as the interaction between flux tubes and convection flows, which may be the origin of bearing flare-productive ARs \citep{2019LRSP...16....3T}. Moreover, realistic dynamo simulations and helioseismology studies will help us better understand the formation and evolution of flaring ARs even before their emergence on the solar surface.

In the last decades, especially since the sequential (vector) magnetograms became available from space, we have deepened our understanding of the usefulness of magnetic helicity. While there still remain difficulties in accurately estimating the magnetic helicity, such as those caused by the aforementioned ambiguity in deriving the transverse magnetic field in the photosphere, we expect that such difficulties can be overcome or at least mitigated in the near future. For example, methods that employ stereoscopic disambiguation may be helpful (e.g., with Solar Orbiter: \citealt{2020A&A...642A...2R,2022SoPh..297...12V}). The photospheric magnetic field, from which the coronal field is reconstructed, deviates significantly from the force-free state. However, it may be possible to improve the coronal field reconstruction by incorporating the chromospheric field, which yields a better force-freeness (e.g., with DKIST and Sunrise-3: \citealt{2008SoPh..247..249W,2019ApJ...870..101F}). Numerical simulations have been used to validate the helicity estimation methods, but by utilizing more realistic flux-emergence simulations that have appeared in recent years, it will be possible to examine these methods with even higher accuracy \citep{2019NatAs...3..160C,2019ApJ...886L..21T,2020MNRAS.494.2523H}. These advanced studies are expected to lead to further understanding and better prediction of flares and CMEs in the near future.

\bibliography{sample}%

\end{document}